\documentstyle[12pt,psfig]{article}
\begin{document}
\raggedbottom 
\begin{center}
\large{Soft Electromagnetic Radiations from Relativistic Heavy Ion
Collisions}
\footnote{Based on the talk presented by Pradip Kumar Roy in ICPA'QGP-1997, 
Jauipur, India.}
\vskip .1in
\vskip .1in
{a) \it Variable Energy Cyclotron Centre, 1/AF Bidhan Nagar, Calcutta,
India}\\
{b) \it Saha Institute of Nuclear Physics, 1/AF Bidhan Nagar, Calcutta,
India}\\
\end{center}
\vskip 0.2cm

\section*{I. Introduction}

    Photons and dileptons have long been considered as excellent probes
of quark-gluon plasma (QGP) expected to be formed in relativistic heavy
ion collisions. However, while evaluating these type of signals, the
immediate interests were to study quasi-particle scattering in dense
matter. In contrast, little has been done to account for the radiative
or bremsstrahlung contributions to photons and dileptons from these
quasi-particles. As will be shown these contributions are dominant as
long as the energy remains below 1 GeV. It has also been argued that
the system formed in relativistic heavy ion collisions undergoes rapid
transverse expansion during the latter stages of the collision. Since
low energy photons/dileptons are produced mostly from the late stage, they
will be affected by the collective transverse flow. Therefore,  one can
get idea about the flow by studying low energy photons and dileptons. 
We apply soft photon approximation (SPA) to evaluate soft photons and
dileptons from quark matter as well as hadronic matter. The estimations
of soft electromagnetic radiations that exist in the literature have
certain discrepancies (see Ref.~\cite{lichard} for details) which we 
have tried to correct in our calculations.

\section*{II. Formulation}

   The emission of soft photons (real or virtual) occurs from external
lines in any process and the probability for such emission is given by
the classical result~\cite{jackson}. The emission of photon from the
interior of the scattering vertex is neglected because in the limit of
very low energy of the emitted photon, its contribution is very low.
The cross-section for the emission of soft real photon produced from
a process $a\,b\,\rightarrow\,c\,d\,\gamma$ is given
by~\cite{drell} is given by
\begin{equation}
q_0\,\frac{d\sigma^{\gamma}}{d^3q} =
\frac{\alpha}{4\,\pi^2}\,\frac{{\hat \sigma(s)}}{q_0^2}
\end{equation}  
and for the emission of a soft dilepton we have
\begin{equation}
\frac{d\sigma^{e^+\,e^-}}{dM^2\,d^2M_T\,dy} =
\frac{\alpha}{12\,\pi^3\,M^2}\,\frac{{\hat \sigma(s)}}{q_0^2}
\end{equation}  
Using kinetic theory we can write down the rate of production
of soft photons from a system at temperature $T$ as
\begin{eqnarray}
E_{\gamma}\,\frac{dN^{\gamma}}{d^4x\,d^3q}&=&
\frac{T^6\,g_{ab}}{16\pi^4}\,\int_{z_{\mathrm {min}}}^{\infty}\,
dz\,\frac{\lambda(z^2T^2,m_a^2,m_b^2)}{T^4}\nonumber\\
&&\,\times\,\Phi(s,s_2,m_a^2,m_b^2)\,K_1(z)\,\left(q_0\,
\frac{d\sigma_{ab}^{\gamma}}{d^3q}\right),
\end{eqnarray}
where $z_{\mathrm {min}} = (m_a+m_b)/T, z = \sqrt{s}/T$, and
$g_{ab} = N_a\,N_b\,(2s_a+1)\,(2s_b+1)$ is the colour and spin
degeneracy appropriate for the collisions.
Similarly we can obtain the rate for soft dilepton
emission.
Here we note that ${\hat \sigma(s)}$ consists of two parts-
electromagnetic and strong or elastic. To evaluate the elastic
cross-section in QGP sector we follow the prescription of
Ref.~\cite{danei}. In the hadronic sector we assume a model
Lagrangian~\cite{haglin} for the calculation of strong cross-section.

Once the rates are obtained one can apply it for an evolving
system to get 
\begin{equation}
\frac{dN}{d^2q_T\,dy}=\int\,\tau\,d\tau\,r\,dr\,d\phi\,d\eta\,
\left[f_Q\, q_0\frac{dN^q}{d^4x\,d^3q}+(1-f_Q)\,
q_0\frac{dN^{\pi}}{d^4x\,d^3q}\right]
\end{equation}
where $f_Q(\tau)$ gives the fraction of the quark-matter\cite{kkmm}
in the system. 

\section*{III. Results}

We calculate the transverse momentum distribution of soft photons and
dileptons for a system undergoing transverse expansion. For this we will
assume that initially a thermalised and chemically equilibrated QGP is 
formed in Pb-Pb collisions at proper time $\tau_i$ = 1
fm/c~\cite{bjorken}. Cooling due to expansion leads to first
order phase transition at T = 160 MeV. When the conversion to hadronic
matter is complete a pure hadron phase is realised
which then freezes-out at T = 140 MeV. The complete dominance
of soft photon multiplicity in the region of 0.1$<p_T$(GeV)$<1.0$
is seen from fig.~(1). 
\begin{figure}
\centerline{\psfig{figure=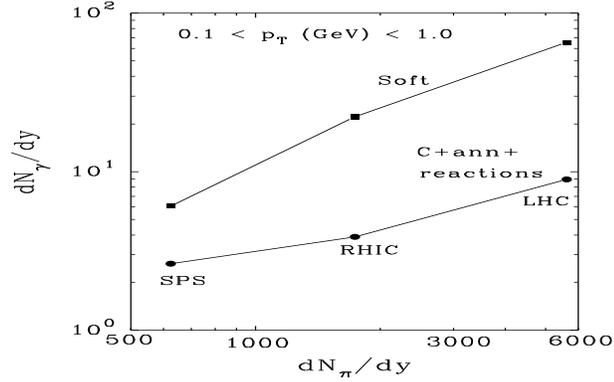,height=5cm,width=8cm}}
\caption{Soft photons vs. photons from Compton plus annihilation
processes from the QGP and hadronic reactions at SPS, RHIC, and LHC
energies from central collision of two lead nuclei.} 
\end{figure}

The result for transverse mass distribution for the low mass dileptons
at RHIC energies is shown in fig.~(2). We see that the pion driven
processes dominate the yeild at lower $M_T$. However at larger $M_T$,
the contributions of quark and pion driven processes are almost same.
This is a reflection of larger temperature in the quark phase, and a
larger effect of transverse flow during the hadronic phase. However,
the invariant mass spectrum does not show up this type of feature.
Thus we see complete dominance of bremsstrahlung dileptons in the
low mass region. Similar charecteristics have been observed at SPS and
LHC energies~\cite{dpal}.

\begin{tabular}{cc}
\psfig{figure=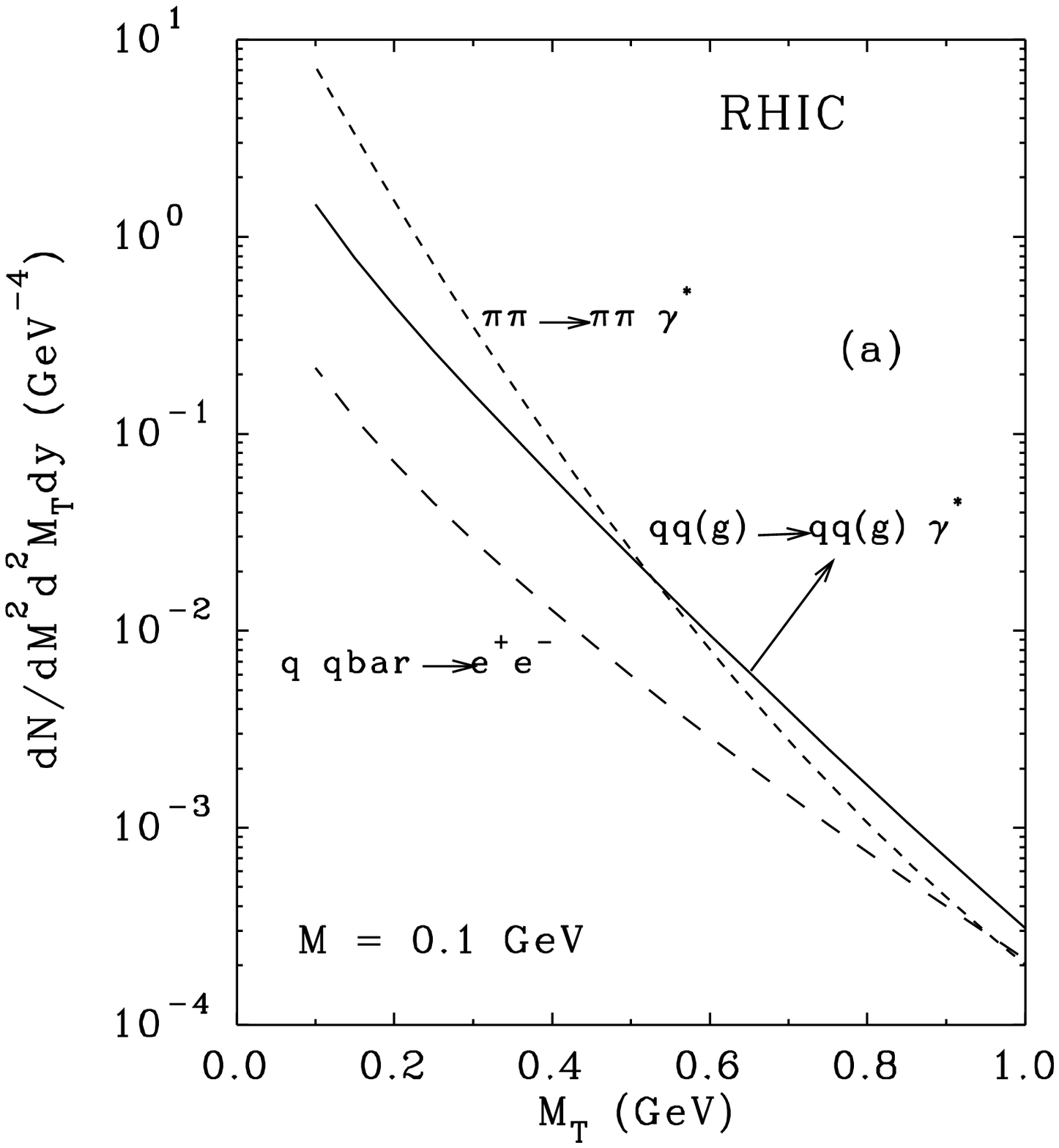,height=6.0cm,width=5.5cm} &
\psfig{figure=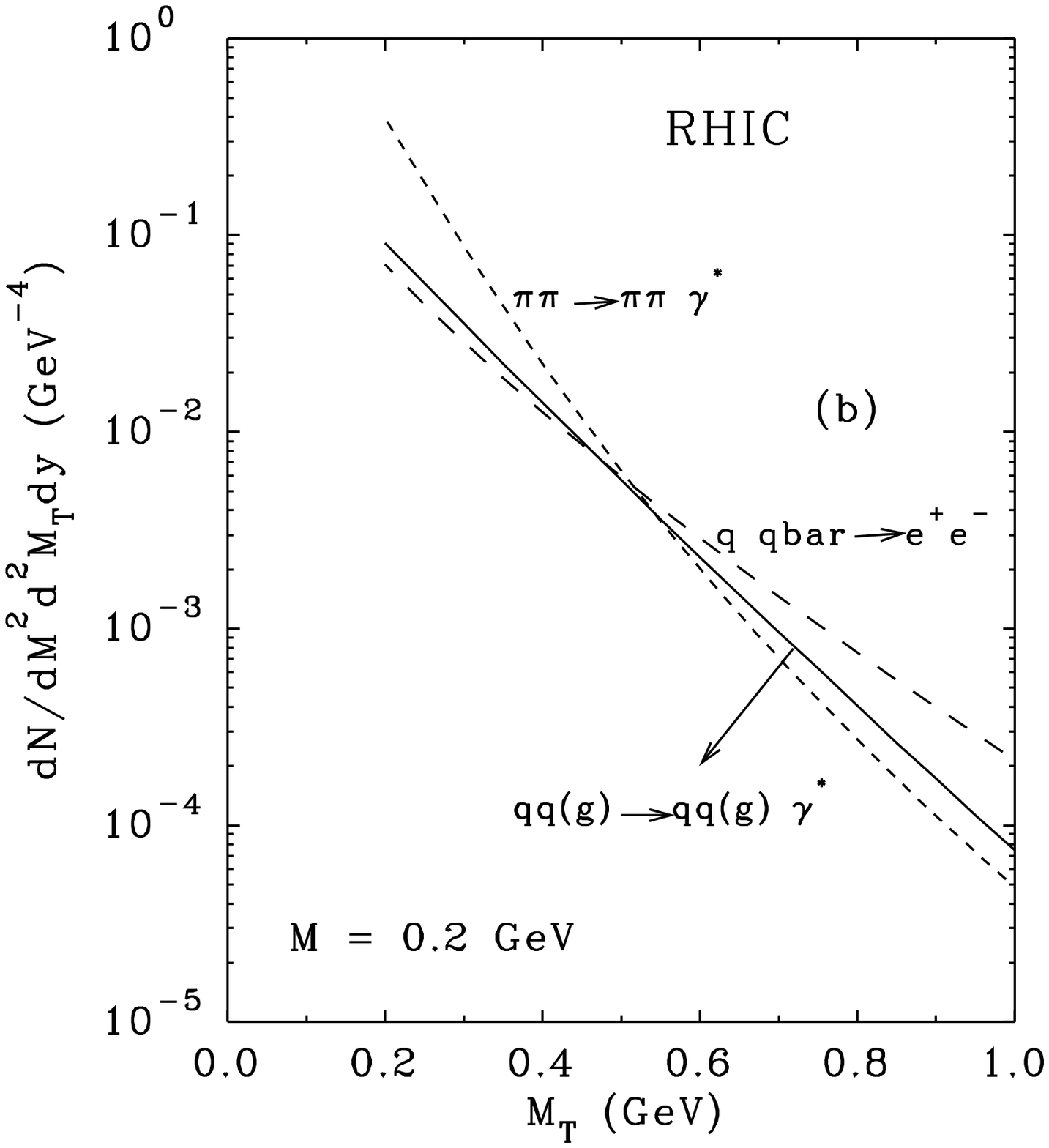,height=6.0cm,width=5.5cm} \\

\psfig{figure=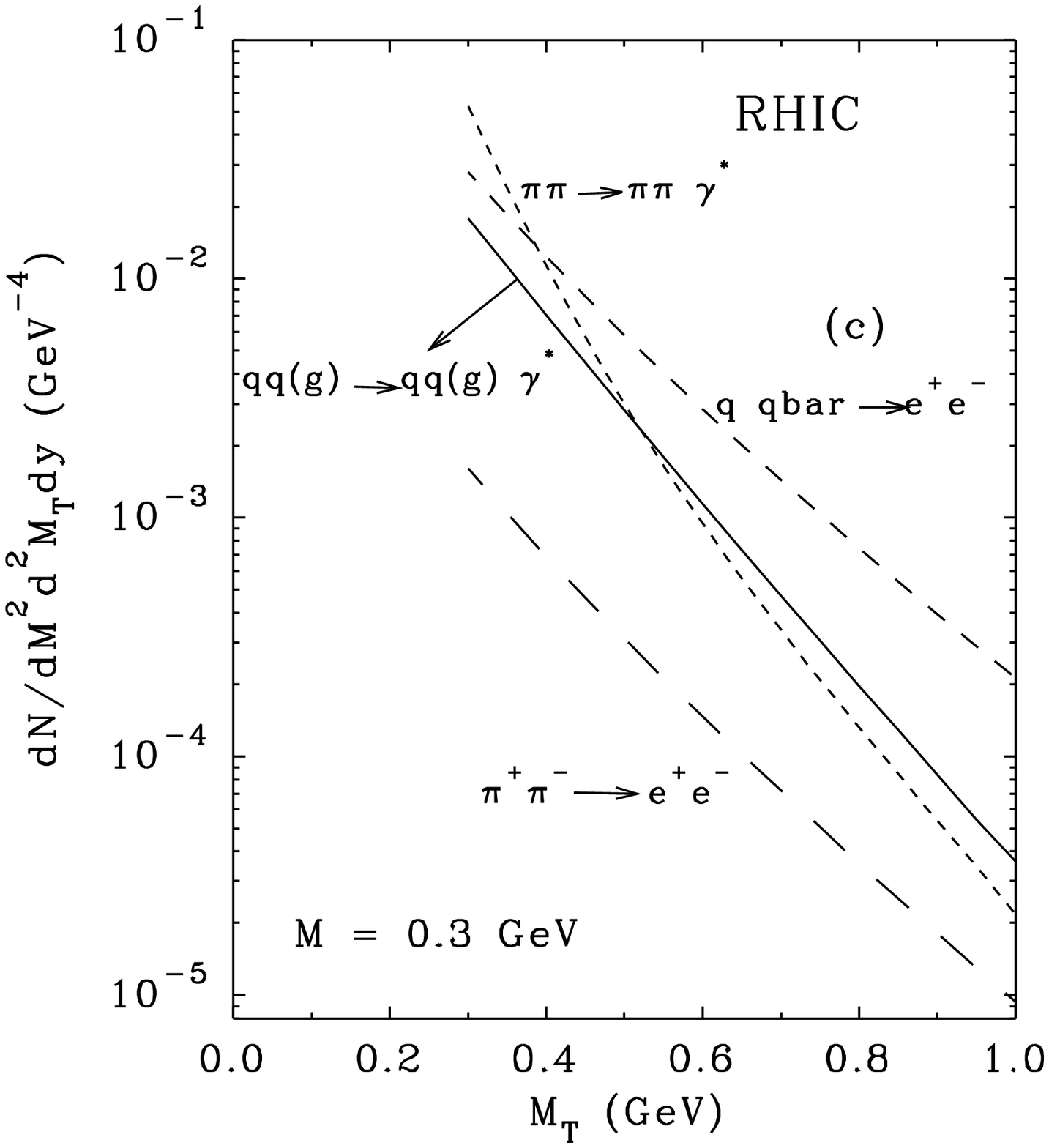,height=6.0cm,width=5.5cm} &
\psfig{figure=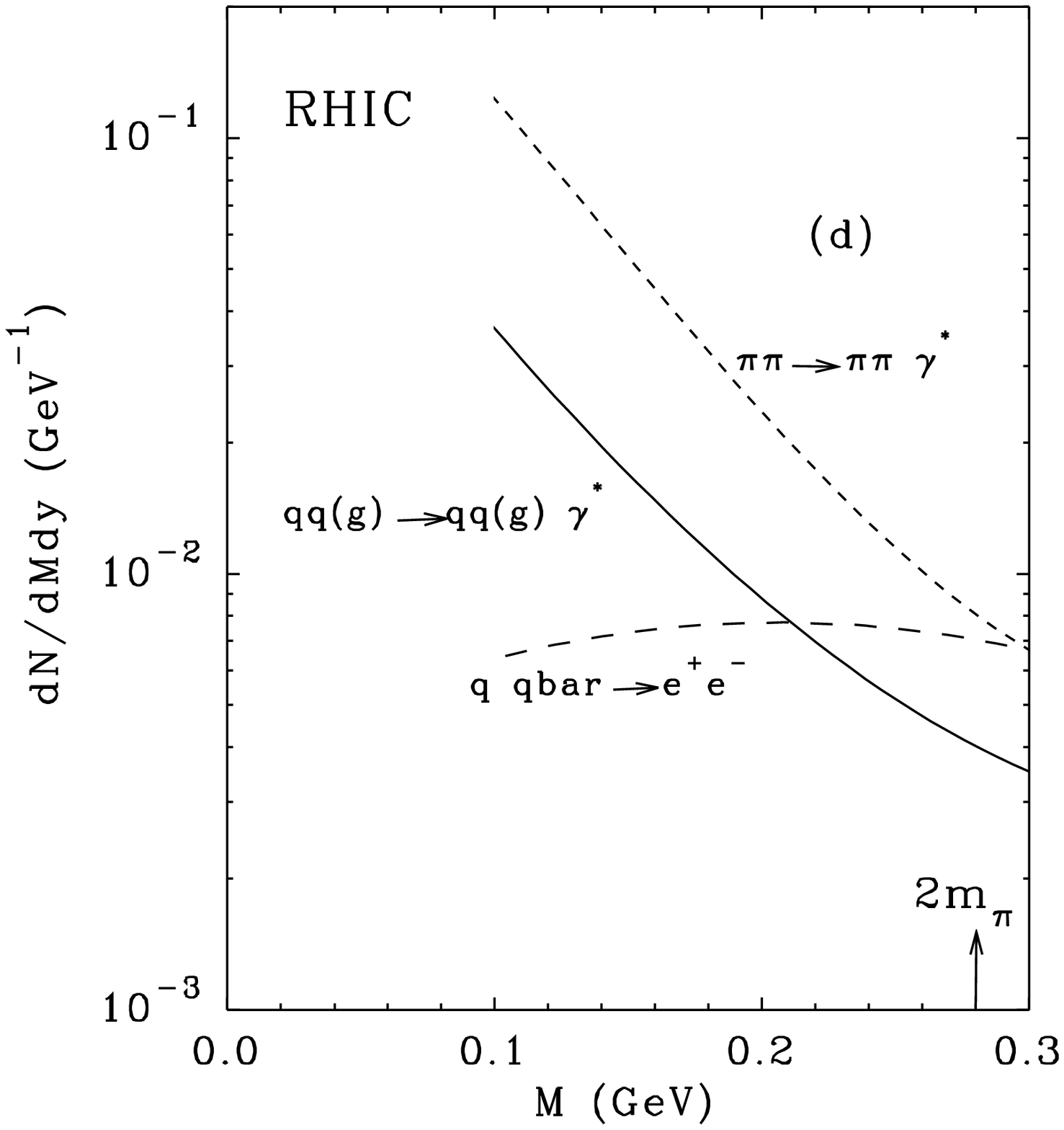,height=6.0cm,width=5.5cm} \\
\end{tabular}

\noindent{
{Figure 2(a--d): The transverse mass distribution of low mass dielectrons
at RHIC energies including bremsstrahlung process and annihilation 
process in the quark matter and the hadronic matter. We give the results
for invariant mass M equal to
 0.1 GeV (a), 0.2 GeV (b), and  0.3 GeV (c) 
respectively. The invariant mass distribution of low mass
dielectrons are also shown (d).}}

\section*{IV. Conclusion}

We have calculated the transverse momentum distributions of photons and
dileptons within a soft photon approximation at SPS, RHIC and LHC
energies. We find that the formation of such a system may be
characterised by an ``intense glow'' of soft electromagnetic radiations
whose feature sensitively depends on the last stage of evolution once we
remove the background of decay photons and dileptons and thus holds out
the promise that soft electromagnetic radiations may be utilised as
chroniclers of final moments of relativistic heavy ion collisions.

  We have kept our discussions limited to photon energies of more than
100 MeV in the hope that Landau-Pomeranchuk-Migdal suppression there may not be
substantial. However, a qualitative argument~\cite{pradip1} shows
that LPM suppression is only marginal in the hadronic sector. Also in 
the quark sector this suppression may not be quite severe, though may 
not be as justified as pion driven processes in the energy region 
considered here.
\vskip .3cm
\noindent
{\bf This talk is an abridged version of Ref.~\cite{dpal}.}

\end{document}